\documentclass[aps,twocolumn,superscriptaddress]{revtex4}
\usepackage{graphicx}

\newcommand{\bra}[1] {\left\langle #1 \right|}
\newcommand{\ket}[1] {\left| #1 \right\rangle}
\newcommand{\prodket[2]} {\left| #1 \right\rangle \otimes
\left| #2 \right\rangle}

\begin{document}

\title{Detecting mode entanglement: The role of coherent states,
superselection rules and particle statistics}

\

\author{S. Ashhab}
\affiliation{Frontier Research System, The Institute of Physical
and Chemical Research (RIKEN), Wako-shi, Saitama 351-0198, Japan}

\author{Koji Maruyama}
\affiliation{Frontier Research System, The Institute of Physical
and Chemical Research (RIKEN), Wako-shi, Saitama 351-0198, Japan}
\affiliation{CREST, Japan Science and Technology Agency,
Kawaguchi, Saitama 332-0012, Japan}

\author{Franco Nori}
\affiliation{Frontier Research System, The Institute of Physical
and Chemical Research (RIKEN), Wako-shi, Saitama 351-0198, Japan}
\affiliation{Physics Department, Michigan Center for Theoretical
Physics, Applied Physics Program, Center for the Study of Complex
Systems, The University of Michigan, Ann Arbor, Michigan
48109-1040, USA}

\date{\today}

\begin{abstract}
We discuss the possibility of observing quantum nonlocality using
the so-called mode entanglement, analyzing the differences between
different types of particles in this context. We first discuss the
role of coherent states in such experiments, and we comment on the
existence of coherent states in nature. The discussion of coherent
states naturally raises questions about the role of particle
statistics in this problem. Although the Pauli exclusion principle
precludes coherent states with a large number of fermionic
particles, we find that a large number of fermionic coherent
states, each containing at most one particle, can be used to
achieve the same effect as a bosonic coherent state for the
purposes of this problem. The discussion of superselection rules
arises naturally in this context, because their applicability to a
given situation prohibits the use of coherent states. This
limitation particularly affects the scenario that we propose for
detecting the mode entanglement of fermionic particles.
\end{abstract}


\maketitle

\section{Introduction}

Entanglement is probably the most intriguing aspect of quantum
mechanics. It has steadily been the subject of research and
controversy ever since it was noted by Schr\"odinger in 1935
\cite{Schroedinger,Einstein,EntanglementReviews}.

The most studied form of entanglement is the one involving two
physical objects with internal degrees of freedom. The
quintessential example in the literature is the entanglement in
the Bell states of two spin-1/2 particles. This simple form of
entanglement, however, is not the only one in nature. In
particular, we consider here mode entanglement, which introduces
additional intrigue to this subject due to the fact that it can
involve the vacuum as a crucial element in the problem, and it can
be obtained using a {\it single} particle. Both of these aspects
are commonly seen as foreign to the discussion of entanglement.

In order to capture the essence of mode entanglement, one can
consider a single particle in a quantum superposition of being at
two different locations. One rarely associates this state with
entanglement. However, if the particle is viewed as an excitation
of an underlying field, the quantum state takes the form of an
entangled state: the first mode of the field containing a particle
while the second mode is empty, and vice versa.

The above example shows that the formal expression used to
describe a quantum state is not the ideal indicator of the
presence of entanglement. Instead, it would be more meaningful to
define the presence of entanglement according to the possibility
of experimentally observing quantum effects associated with
entanglement, e.g.~the violation of the Bell inequalities
\cite{Bell}. As we shall discuss in some detail below, the nature
of the particles involved in the mode entanglement is a crucial
factor in determining whether this entanglement is detectable or
not. Analyzing the detectability of mode entanglement for
different types of particles is the main subject of this paper.

Several theoretical studies have analyzed the so-called
single-photon entanglement in quite some detail
\cite{Tan,Hardy,Wiseman,Hessmo,vanEnk,Bartlett1}. In fact, there
have been experimental tests of the Bell inequalities using the
mode entanglement of single photons \cite{Babichev,DAngelo}.
Photons, however, represent a single type of particles with
specific properties. Here we build on the results of
Ref.~\cite{Ashhab} (see also Refs.~\cite{Aharonov,TerraCunha}): we
analyze the roles played by particle statistics \cite{Omar} and
superselection rules \cite{Wick} in the detectability of mode
entanglement. We divide our discussion into four cases, depending
on the nature of the particles, i.e.~bosons or fermions, and
whether superselection rules constrain the total particle number
to be fixed or not. This division simplifies the task of
identifying the roles played by the different physical elements in
the problem.

The importance of superselection rules, i.e.~the constraint of
having a fixed particle number, can be seen by considering a
Bell-violation experiment. In such an experiment, it is necessary
to perform measurements in a variety of bases. In the case of mode
entanglement, the notion of measurements in different bases
suggests that one needs to perform measurements in bases of
indefinite particle number, e.g.~the basis
$(\ket{0}\pm\ket{1})/\sqrt{2}$. If superselection rules apply to
the type of particle under consideration, such a measurement is
forbidden. Although this difficulty might seem to be a major
obstacle to the detectability of mode entanglement under the
constraint of superselection rules, we shall present procedures to
overcome it by utilizing the indistinguishability between the
particle under consideration and other properly prepared ancillary
particles. The use of particle indistinguishability in our
proposed procedures indicates that particle statistics will also
be an important factor in the detectability of mode entanglement,
since there are major differences between bosonic and fermionic
particles in this regard.

The role of coherent states in a Bell-test experiment can also be
seen by considering the need for performing measurements in a
variety of bases. Rotations on a quantum state before the
measurement are equivalent to, and sometimes necessary for,
changing the measurement basis. Such rotations are commonly
induced using coherent states. Thus, in our analysis below we
shall deal with questions related to coherent states,
superselection rules and particle statistics.

This paper is organized as follows: In Sec.~II we present the
basic setup for our analysis. We analyze
mode-entanglement-detection (gedanken) experiments using four
different types of particles in Secs.~III-VI. We conclude by
reviewing our main results in Sec.~VII.

\section{Description of the setup}

Throughout this paper we consider a setup where a particle is
prepared in a spatially delocalized state of the form:
\begin{equation}
\ket{\Phi} = \frac{1}{\sqrt{2}} \left( \ket{L} + \ket{R} \right),
\label{eq:Psi_after_BS}
\end{equation}
where the states $\ket{L}$ and $\ket{R}$ are thought of as being
localized on opposite sides of the experimental setup. For the
case of photons, for example, this state can be obtained by
sending a beam into a 50/50 beam splitter. When viewed as a state
of the photon field, in the form
\begin{equation}
\ket{\Phi} = \frac{1}{\sqrt{2}} \left( \ket{10} + \ket{01}
\right),
\label{eq:Psi_after_BS_Fields}
\end{equation}
one can see that this is an entangled state. The task is now to
detect this entanglement, e.g.~using a Bell-test experiment
\cite{Bell}.

How to proceed in order to probe the entanglement in the state in
Eq.~(\ref{eq:Psi_after_BS_Fields}) depends on the available
measurement tools. For example, the experiments on this subject
\cite{Babichev,DAngelo} probed the entanglement using homodyne
detection, mixing the incoming photons with coherent states of
known phases. Analyzing the detailed description of such
specialized techniques, however, would be a distraction from the
aim of this paper. We therefore consider a conceptually simpler
scenario: we imagine that the incoming (flying) particle can
excite a two-level target particle from its ground state to its
excited state. The initial state of the combined system is given
by
\begin{equation}
\ket{\Psi} = \frac{1}{\sqrt{2}} \left( \ket{10} + \ket{01} \right)
\otimes \ket{gg},
\end{equation}
where the first ket describes the state of the flying particle,
the second ket describes the state of the two target particles
(note that one target particle is placed on each side of the
setup), and the symbols $g$ and $e$ are used to denote the ground
and excited states of the target particles. Depending on whether
the flying particle is absorbed by the target particle during the
excitation process or not, one obtains either the state
\begin{equation}
\ket{\Psi} = \frac{1}{\sqrt{2}} \ket{00} \otimes \left( \ket{eg} +
\ket{ge} \right)
\label{eq:Psi_after_absorption}
\end{equation}
or the state \cite{ExcitationProcess}
\begin{equation}
\ket{\Psi} = \frac{1}{\sqrt{2}} \left( \prodket[10]{eg} +
\prodket[01]{ge} \right).
\label{eq:Psi_after_excitation}
\end{equation}
The state in Eq.~(\ref{eq:Psi_after_absorption}) is the proper
description for an incoming photon that is absorbed by one of two
target atoms. However, it cannot be obtained whenever
superselection rules apply to the species of flying particles,
since the flying particle cannot be annihilated in this case. We
shall refer to particles with a conserved total number as massive
particles (more as a matter of easily recognizable terminology
than fundamental physical arguments \cite{MassArgument}).

Note that the number of target particles does not change in the
above picture, and they do not move between the two sides of the
experimental setup. The discussion of superselection rules is
therefore not crucial in regard to the target particles. It is
safest, however, to assume that they are different from the
flying-particle species, such that we do not need to worry about
complications associated with the flying and target particles
being indistinguishable and obeying identical-particle symmetry
constraints.

\section{Case 1: Massless bosons}

We start by considering the relatively simple case of a single
photon passing through a beam splitter and resulting in an
entangled state between the left and right modes of the
electromagnetic field. Although the experiments of
Refs.~\cite{Babichev,DAngelo} relied on an auxiliary laser beam as
a reference phase standard for the Bell test, it seems
conceptually simpler to imagine the incoming photons being
absorbed by target atoms and resulting in states of the form given
in Eq.~(\ref{eq:Psi_after_absorption}). The mode entanglement is
then transferred to the internal degrees of freedom of the target
atoms. One can then conclude that the measurements for the Bell
test can be performed straightforwardly on the states of the
target atoms.

An important point that was not addressed in the above scenario is
the fact that for a Bell test one would need to perform rotations
on the states of the target atoms before the measurement (here we
are making the realistic assumption that measurements will always
be performed in the $\{\ket{g},\ket{e}\}$ basis). Such rotations
are typically performed using classical fields of the same
frequency as the incoming photons. For these fields to be
classical and useful for our purposes, one must know the relative
phase between the fields on the left and right sides of the beam
splitter. In other words, although the entanglement was
transferred to the internal states of the target atoms when the
incoming photon was absorbed, one still needs to have a common
phase reference (typically in the form of photonic coherent states
on the two sides of the experimental setup with a known relative
phase). The simple-looking scenario of using target atoms
therefore does not eliminate the need for a phase reference. It
only divides the procedure into two steps, each of which is
conceptually simple.

Having established the need for a common phase reference, we are
now led to ask whether the two sides of the setup must be
entangled in order to have such a common phase reference. If the
answer is yes, one would be led to question whether any observed
phenomena probed the mode entanglement of the incoming photons or
a combination of the mode entanglement and the pre-existing
entanglement in the setup. We address the above question next.

\subsection{Can two coherent states with a known relative phase be
prepared independently of each other?}

Although the answer to the above question, in the affirmative, is
accepted by the majority of physicists, it has generated some
controversy in recent years \cite{Molmer,Bartlett2}. We therefore
address it explicitly here for clarity.

The simplest approach to take here is probably to consider the
classical problem of, say, radio-frequency antennas. Taking two
distant antennas with known relative orientations, and assuming
the antennas are controlled by experimentalists with synchronized
clocks, the two experimentalists can produce classical waves with
a known relative phase. If the setup includes a screen, i.e.~a set
of detectors, one can predict exactly where the interference
maxima and minima will appear on the screen. All that is needed to
make this prediction is knowledge of the relative orientation of
the antennas and synchronization of the clocks. Although this
argument treats relatively low-frequency waves, there is
conceptually nothing different when dealing with the optical
frequencies. Finally, when this situation is described in
quantum-mechanical terms, the predictability of the interference
patterns implies that the photon states generated by the two
sources must be coherent states.

One can therefore conclude that as long as the two sources share
reference frames and synchronized clocks, they can in principle
generate coherent states with a known relative phase. The fact
that present-day experiments cannot produce two independent
optical-frequency lasers with a known relative phase should not be
seen as a fundamental obstacle to the existence of coherent states
(as was in fact noted in Ref.~\cite{Molmer}). The most crucial
point here is probably the fact that the two sources generating
the mutually coherent waves do not need to share any entanglement.

Turning back to the problem of performing rotations on an atomic
state, one can also envision replacing the common phase reference
by the application of intense static electric fields (the strength
of the field being compared with the frequency of the relevant
atomic transition) in order to perform the atomic-state rotation.
The phase-standard aspect of the shared reference frame disappears
completely in this case. One must keep in mind, of course, that
real atoms cannot be approximated by two-level systems under such
intense fields. However, this argument demonstrates that sharing a
common phase reference is nothing more than sharing a space and
time reference frame.

As for the need to share reference frames, this is by no means
unique to the case of quantum-optical coherent states. It also
applies, e.g., to a Bell-test experiment using spin states. More
specifically, take two observers that share maximally entangled
pairs of spin-1/2 particles (e.g.~in the singlet state). Until the
observers establish the proper reference frames for their
measurements, they cannot detect the entanglement. Of course they
can sacrifice a few pairs in order to establish those proper
reference frames, and then they can proceed with the experiment
and observe the violation of the Bell inequality. Alternatively,
the two observers can scan the entire range of possible
measurement directions, thus simultaneously establishing the
common reference frame and observing the Bell-inequality
violation. The main point here, however, is to note that in many
(classical and quantum) physical problems a common reference frame
must be established before correct predictions can be made.

\section{Case 2: Massive bosons}

This case was analyzed in Ref.~\cite{Ashhab}, and we shall not
repeat the analysis here. The main result is that if one takes $N$
ancillary particles of the same species as the flying particles
and forms two entangled Bose-Einstein condensates (in a properly
prepared state), one can follow the procedure explained in
Ref.~\cite{Ashhab} and detect the mode entanglement in the state
of the flying particles. The observable concurrence for each
incoming flying particle is given by $1-1/(2N)$ for large $N$.

An important result in this case is that the condensate of $N$
particles can be reused for an arbitrary number of flying
particles. The unlimited reusability of the condensate suggests
that the condensate can be naturally thought of as playing an
auxiliary role in the experiment. This result is also rather
counterintuitive, and it stands in contrast with the notion that
quantum reference frames are generally degraded as a result of
repeated use \cite{Bartlett3}. A possible explanation of this
result is that in the procedure of Ref.~\cite{Ashhab} no
measurements are performed directly on the condensate. In fact, if
one performs measurements on the condensate, one can (at least
probabilistically) increase the entanglement in the first created
pairs of target particles, but the entanglement of subsequent
pairs will be degraded. It would be interesting to see if similar
ideas can be applied to quantum reference frames in general.

It should be noted here that as the flying particles come into the
proposed setup and are used to excite the target particles then
properly discarded into the condensate, some amount of
entanglement between the condensate and the target particles is
generated. The state of the condensate therefore changes after
each measurement on a given pair of target particles.
Alternatively, if several entangled pairs are generated before any
measurement is performed, the different pairs will be entangled
with each other. As such, the different entangled pairs generated
in this procedures cannot be considered independent and
identically distributed (i.i.d.). Note, however, that whenever the
Bell inequalities are violated, the observed correlations cannot
be described by local-hidden-variable theories. In other words,
i.i.d.-ness of the source is not a requirement of the Bell test.

In principle, it is possible to write down the full (pure) quantum
state of the entire system and analyze the entanglement present in
different sets of subsystems. However, since our main focus in
this paper is the detection of mode entanglement, we only consider
the correlations that are present within the individual pairs of
target particles, even in the case where a stream of flying
particles is used to generate a large number of entangled pairs of
target particles. Other correlations in the system give rise to
interesting phenomena that are not directly related to the aim of
this paper and will be discussed in more detail elsewhere.

Another interesting result in the case of massive bosons is that a
single ancillary particle is sufficient to allow the observation
of the Bell-inequality violation (ensemble averaging over many
setups is needed in order to guarantee the violation, as will be
discussed in detail elsewhere). This result can be verified by
using the following criterion presented in Ref.~\cite{Horodecki}.
First, following Ref.~\cite{Ashhab} with $N$ ancillary particles,
we calculate the reduced density matrix describing the state of
the target particles in the basis
$\{\ket{gg},\ket{ge},\ket{eg},\ket{ee}\}$, and we find it to be
given by:
\begin{equation}
\rho_{\rm TP} = \frac{1}{2} \left(
\begin{array}{cccc}
0 & 0 & 0 & 0 \\
0 & 1 & \gamma & 0 \\
0 & \gamma & 1 & 0 \\
0 & 0 & 0 & 0 \\
\end{array}
\right),
\end{equation}
where $\gamma \approx 1-1/(2N)$ [In the following we only need to
use the fact that $\gamma$ is nonzero]. Using this density matrix,
we now follow Ref.~\cite{Horodecki} and define a $3\times 3$
matrix $T$ with entries $T_{ij} \equiv {\rm Tr} \left[ \rho_{\rm
TP} \left( \sigma_i^{\rm L} \otimes \sigma_j^{\rm R} \right)
\right]$ with the standard Pauli matrices $\sigma_1$, $\sigma_2$,
and $\sigma_3$ for the left and right particles. Then we compute
the three eigenvalues of the matrix $T^\dagger T$ and define a new
function $M(\rho)$ as the sum of the two greatest eigenvalues. The
necessary and sufficient condition for the violation of the Bell
inequality (in the Clauser-Horne-Shimony-Holt version
\cite{Clauser}) can be expressed as $M(\rho)>1$. For the density
matrix $\rho_{\rm TP}$ above, we find that $M(\rho) = 1+|\gamma|^2
\approx 1+[1-1/(2N)]^2$, which is always greater than 1 regardless
of $N$, hence the violation of the Bell inequality.

\section{Case 3: Massless fermions}

We now turn to the case of fermionic flying particles. We start by
considering the case of massless fermions because it gives
conceptually interesting results and serves as an introduction to
Sec.~VI, regardless of whether it corresponds to any realistic
physical situation. The discussion would also be relevant if
superselection rules do not have to be obeyed for fermionic
particles, a situation predicted by some high-energy theories
\cite{Aharonov}.

We consider a (possibly hypothetical) fermionic analog of photons:
we imagine a fermionic species of particles that can be created at
will, and any given mode can contain at most one particle. We
therefore cannot create coherent states of a form similar to
coherent states of bosonic particles, i.e.
\begin{equation}
\ket{\psi}_{\rm coherent,B} = \exp
\left\{-\frac{|\eta|^2}{2}\right\} \sum_{n=0}^{\infty}
\frac{\eta^n}{\sqrt{n!}} \ket{n}.
\end{equation}
We shall show, however, that the fermionic analogue of coherent
states can be used to achieve the same result obtained using
bosonic coherent states in the context of the present discussion.
As mentioned above, we assume that states of the form
\begin{equation}
\ket{\psi}_{\rm coherent,F} =  \frac{1}{\sqrt{2}} \left( \ket{0} +
\ket{1} \right)
\label{eq:FermiCohState}
\end{equation}
are physical and can be created at will. The above state will be
the main building block for the coherent-state-like manipulations
below.

We now imagine that the incoming particle is absorbed by one of
two target particles as explained in Sec.~II. This can be achieved
using the effective Hamiltonian:
\begin{equation}
\hat{H} = J \left( i \sigma_+ a - i \sigma_- a^{\dagger} \right),
\label{eq:JCHamiltonian}
\end{equation}
where $J$ is the coupling strength, $\sigma_{\pm}$ are raising and
lowering operators of the target-particle state
($\sigma_+\ket{g}=\ket{e}$), and $a$ and $a^{\dagger}$ are,
respectively, annihilation and creation operators of the incoming
particle species. After the absorption of the flying particle, the
target particles end up in a state of the form given in
Eq.~(\ref{eq:Psi_after_absorption}).

As discussed in Sec.~III above, the detection of mode entanglement
is now reduced to the ability of performing arbitrary rotations on
the states of the target particles. We therefore focus on these
rotations for the remainder of this section, and below we give
explicit expressions for the representative example of a $\pi/2$
rotation. Note that we do not allow using a bosonic coherent state
here; instead we imagine that the target particle can only be
manipulated using the same Hamiltonian describing the absorption
of the incoming particle (Eq.~\ref{eq:JCHamiltonian}).

Let us take a target particle in an arbitrary initial state
\begin{equation}
\ket{\psi}_i = \alpha \ket{g} + \beta \ket{e}
\label{eq:Psi_before_rotation}
\end{equation}
and try to rotate it to the state
\begin{equation}
\ket{\psi}_{f, \rm ideal} = \frac{\alpha - \beta}{\sqrt{2}}
\ket{g} + \frac{\alpha + \beta}{\sqrt{2}} \ket{e}.
\end{equation}
The above quantum state can also be described using the density
matrix
\begin{equation}
\rho_{f, \rm ideal} = \frac{1}{2} \left(
\begin{array}{cc}
|\alpha-\beta|^2 & (\alpha+\beta)^* (\alpha-\beta) \\
(\alpha+\beta) (\alpha-\beta)^* & |\alpha+\beta|^2
\end{array}
\right).
\label{eq:rhoIdeal}
\end{equation}
In order to perform the desired rotation, one can try to use an
ancillary mode in the state given by Eq.~(\ref{eq:FermiCohState})
and allow that mode to interact with the target particle using the
effective Hamiltonian in Eq.~(\ref{eq:JCHamiltonian}) for a
duration of $\pi/(4J)$. If we trace out the degrees of freedom of
the ancillary mode at the final time, we find that the above
operation transforms the initial state of the target particle
(Eq.~\ref{eq:Psi_before_rotation}) into a mixed state described by
the density matrix:
\begin{widetext}
\begin{equation}
\rho_f = \frac{1}{4} \left(
\begin{array}{cc}
2 |\alpha|^2 + |\alpha-\beta|^2 & \sqrt{2} \alpha (\alpha+\beta)^*
+ \sqrt{2} (\alpha-\beta) \beta^* \\
\sqrt{2} \alpha^* (\alpha+\beta) + \sqrt{2} (\alpha-\beta)^* \beta
& |\alpha+\beta|^2 + 2 \beta^2
\end{array}
\right).
\label{eq:rho}
\end{equation}
\end{widetext}
The overlap between this state and the ideal state can be
calculated using the fidelity
\begin{equation}
F = {}_{f, \rm ideal} \bra{\psi} \rho_f \ket{\psi}_{f, \rm ideal}.
\end{equation}
We do not write down the long expression for the fidelity in the
above example or go further into specific averaging procedures.
The main point to note is that the fidelity is clearly smaller
than 1 (compare Eqs.~\ref{eq:rhoIdeal} and \ref{eq:rho})
\cite{OptimizedAncillary}.

Since the fidelity reduction can be attributed to instances where
the initial state of the target particle and ancillary mode is
given by $\prodket[g]{0}$ or $\prodket[e]{1}$
\cite{FidelityReduction}, we now try to reduce the impact of such
instances. An obvious approach is to use a large number $N$ of
ancillary modes, each in a state of the form given by
Eq.~(\ref{eq:FermiCohState}); the `bad' states
$\prodket[g]{00...0}$ and $\prodket[e]{11...1}$ now have very
small probability amplitudes. We now perform a numerical
simulation: we take the target particle and allow it to interact
with each ancillary mode using the Hamiltonian in
Eq.~(\ref{eq:JCHamiltonian}) for a duration of $\pi/(4JN)$.
Without going into the details of the calculation, which parallels
the explanation given above for a single ancillary mode, we find
that the fidelity, i.e.~the overlap between the ideal and actual
final states, of the target particle approaches 1, with error
proportional to $1/N$.

The above procedure can therefore be incorporated into a
mode-entanglement experiment, with the conclusion that after the
absorption of the incoming particle an arbitrary measurement can
be performed on the states of the target particles. This result
implies that the mode entanglement would be detectable in a
Bell-test experiment.

We should stress here that the coupling between the target
particle and the ancillary modes must be done sequentially. If,
instead, the target particle is coupled to all ancillary modes
simultaneously using the Hamiltonian
\begin{eqnarray}
\hat{H} & = & J \sum_k \left( i \sigma_+ a_k - i \sigma_-
a_k^{\dagger} \right)
\nonumber \\
& = & J \sqrt{N} \left( i \sigma_+ \sum_k \frac{a_k}{\sqrt{N}} - i
\sigma_- \sum_k \frac{a_k^{\dagger}}{\sqrt{N}} \right),
\label{eq:JCHamiltonian2}
\end{eqnarray}
the target particle couples to a single collective mode, defined
by the annihilation operator $b\equiv \sum_k a_k/\sqrt{N}$. Using
this procedure therefore gives the same results as using a single
ancillary mode, i.e.~a 50\% success probability for producing an
entangled pair of target particles \cite{Ashhab} (here $k$ labels
the different ancillary modes).

\section{Case 4: Massive fermions}

Encouraged by the success achieved using fermionic coherent states
in Sec.~V, we now try to follow a similar procedure for the case
of massive fermions.

Since we now want to impose superselection rules (e.g., unlike the
scenario of Sec.~V, the flying particle is not absorbed upon
exciting the target particle and we cannot create coherent states
at will), we must look for alternatives with a fixed particle
number for the flying-particle species. We follow a procedure
similar to that introduced in Ref.~\cite{Ashhab} and combine it
with the sequential manipulation of Sec.~V.

Our starting point is the initial state of the flying particle and
two target particles given in Eq.~(\ref{eq:Psi_after_excitation}).
We also assume that we have already created $N$ entangled pairs of
ancillary modes (with each pair of modes sharing one particle) of
the form
\begin{equation}
\ket{\Psi_{\rm anc}} = \frac{1}{\sqrt{2}} \left( \ket{L_{\rm anc}}
+ \ket{R_{\rm anc}} \right),
\label{eq:ConservedFermiAncState}
\end{equation}
where the states $\ket{L_{\rm anc}}$ and $\ket{R_{\rm anc}}$
describe the ancillary particle being localized on the left and
right side of the beam splitter, respectively. We now want to
perform a sequence of local operations, each involving a target
particle (on the left or right side), the corresponding
flying-particle mode and an ancillary mode. We shall try to design
this sequence of operations such that the flying particle is
`discarded' into one of the ancillary modes by the end of the
entire procedure (the key property of this `disposal' process is
that one should no longer be able to deduce the location of the
excited target particle from the state of the flying-particle
species). The concurrence in the state of the target particles at
the end of the sequence of operations can be calculated from the
target-particle reduced density matrix, which is obtained by
tracing over the degrees of freedom of the flying and ancillary
particles at the end of the procedure.

We now focus on a single operation to be performed on one side of
the setup; this operation will essentially be the building block
from which the entire sequence is constructed. We look for a
unitary operation that mixes the states
$\ket{e}\otimes\ket{1}_{\rm flying}\otimes\ket{0}_{\rm anc}$ and
$\ket{e}\otimes\ket{0}_{\rm flying}\otimes\ket{1}_{\rm anc}$ with
some probability \cite{OperationNotes}. The desired effect of this
operation is that, if the flying particle is on the side of the
setup where the operation is performed and the ancillary mode is
empty, the flying particle will (with some probability) be
discarded into the ancillary mode, thus partially erasing the
information in the flying-particle mode. In Ref.~\cite{Ashhab}, a
well-merging process was proposed for this purpose. We find the
well-merging process unsuitable for generalization to the
multi-step procedure that we are trying to construct here. It
seems that the next closest analogue to what was done in Sec.~V is
to use operations of the form
\begin{equation}
U = \left(
\begin{array}{cc}
\cos\theta & -\sin\theta \\
\sin\theta & \cos\theta
\end{array}
\right)
\end{equation}
in the above basis (i.e.~$\{\ket{e}\otimes\ket{1}_{\rm
flying}\otimes\ket{0}_{\rm anc},\ket{e}\otimes\ket{0}_{\rm
flying}\otimes\ket{1}_{\rm anc}\}$), while not affecting any other
state \cite{NoFundamentalDifficulty}.

In the case of a single pair of ancillary modes
(Eq.~\ref{eq:ConservedFermiAncState}), the optimal value of
$\theta$ is $\pi/2$ (this would be referred to as a $\pi$
rotation), resulting in a concurrence of $1/2$ between the two
target particles. In this case one can clearly identify the
successful instances as those associated with the subspace
$\{\ket{eg}\otimes\ket{L_{\rm flying}}\otimes\ket{R_{\rm
anc}},\ket{ge}\otimes\ket{R_{\rm flying}}\otimes\ket{L_{\rm
anc}}\}$ and the unsuccessful instances as those associated with
the subspace $\{\ket{eg}\otimes\ket{L_{\rm
flying}}\otimes\ket{L_{\rm anc}},\ket{ge}\otimes\ket{R_{\rm
flying}}\otimes\ket{R_{\rm anc}}\}$. In particular, if both the
flying and ancillary particles end up on the same side of the
setup, their indistinguishability cannot be utilized to `erase'
the information about the location of the flying particle.

We now numerically simulate the procedure with two pairs of
ancillary modes (i.e.~two ancillary particles) and search for the
optimal values of $\theta_1$ and $\theta_2$, which represent the
two steps in the procedure (we take the same value of $\theta_j$
on both sides of the experimental setup in each step). We find
that the the maximum achievable concurrence is still given by
$1/2$, and is obtained by taking one of the two angles equal to
$0$ and the other equal to $\pi/2$. This means that the optimal
approach is to use only one of the two available pairs of
ancillary modes.

Although the above is one specific example of a procedure
attempting to increase the concurrence between the target
particles, it seems to be the most natural one combining the
results of Sec.~V and those of Ref.~\cite{Ashhab}. We therefore
suspect that no other procedure would allow an increase in the
concurrence.

Note that the failure to increase the concurrence using a larger
number of ancillary particles does not mean that quantum-nonlocal
effects cannot be observed in this system. In principle, they are
observable \cite{Ashhab}. The only concern is that one can raise
questions about whether the observed effects should be attributed
to the mode entanglement or the combination of the mode
entanglement and the entanglement already present within the pair
of ancillary modes.

\section{Conclusion}

In this paper we have analyzed the problem of detecting mode
entanglement using various types of particles. The results are
summarized in Table I, assuming the existence of $N$ suitably
prepared ancillary particles for the case of massive particles.
For massless particles, mode entanglement is no different from the
Bell-state entanglement in terms of experimental observability,
regardless of particle statistics. As we have discussed, coherent
states play an important role in this context (note that coherent
states can only be used when considering massless particles). For
massive particles, i.e.~those that must obey particle-number
superselection rules, one must make use of additional ancillary
particles in order to experimentally detect the mode entanglement.
For bosons, an ancillary Bose-Einstein condensate of $N$ particles
can be reused arbitrarily many times, which suggests that the
condensate should be thought of as a catalyst in the experimental
detection of the entanglement. For fermions, we cannot find any
procedure that gives better results than using a single ancillary
particle. This result suggests that one cannot detect the mode
entanglement in this case; one can only detect the entanglement
present in the combination of the flying and ancillary particles.
If this conclusion is correct, one would have to question whether
the mode entanglement of massive fermions can be considered a true
(i.e.~experimentally observable) form of entanglement.

\begin{table}
\begin{tabular}{|l|c|c|} \hline
Particle type     & Concurrence & Max. number \\
                  & between TPs & of repetitions \\
\hline
Massless bosons   & 1           & $\infty$ \\
Massive bosons    & $1-1/(2N)$  & $\infty$ \\
Massless fermions & 1           & $\infty$ \\
Massive fermions  & 1/2         & $N$ \\
\hline
\end{tabular}
\newline
Table I: Concurrence between the two target particles (TPs) for
one incoming flying particle and maximum number of times the
experiment can be repeated (with a given number $N$ of ancillary
particles in the case of massive particles) for different types of
particles.
\end{table}

Finally, we would like to mention that the concept of coherent
fermionic states has been used in the literature \cite{Cahill},
mainly as a simple calculational tool to analyze the behaviour of
fermionic many-body systems (This effort was motivated by the fact
that coherent states provide invaluable predictive power when
studying certain aspects of the behaviour of bosonic many-body
systems \cite{Castin}). In this paper, we have analyzed the
possibility of using fermionic coherent states to simulate
classical fields for the purpose of inducing unitary
transformations on the states of the target particles. The
sequentiality in our proposed procedure provides some
distinguishability between the particles, thus we do not have to
deal with anticommutation rules or Grassmann variables. It would
be interesting to see if there is a connection between the ability
to utilize fermionic coherent states analyzed here and the
properties of these states analyzed in previous work.

\begin{acknowledgments}
This work was supported in part by the National Security Agency
(NSA), the Army Research Office (ARO), the Laboratory for Physical
Sciences (LPS), the National Science Foundation (NSF) grant
No.~EIA-0130383, the Japan Science and Technology Agency (JST) and
the Japan Society for the Promotion of Science Core-to-Core
(JSPS-CTC) program. One of us (S.A.) was supported by the Japan
Society for the Promotion of Science (JSPS).
\end{acknowledgments}


\begin{thebibliography}{99}
\bibitem{ExcitationProcess} One can imagine obtaining the state in
Eq.~(\ref{eq:Psi_after_excitation}) by, e.g., transferring an
internal excitation from the flying to the target particle or
using an auxiliary field that becomes resonant with the target
particle only in the presence of the flying particle. Although
these processes are challenging for present-day experiments, the
important thing to us is that they are consistent with realistic
physical processes.
\bibitem{MassArgument} For example, it was argued in
Ref.~\cite{Aharonov} that superselection rules could, in
principle, be only approximate even for massive particles.
\bibitem{OptimizedAncillary} It can be shown that Eq.~(\ref{eq:FermiCohState})
is the optimal state that maximizes the fidelity, regardless of
the rotation angle (for a $\pi/2$ rotation it gives $F=0.819$).
However, we shall not include the details of the calculation here.
\bibitem{FidelityReduction} This statement is clearer in the
context of the mode-entanglement detection procedure; see
Ref.~\cite{Ashhab}.
\bibitem{OperationNotes} Note that we are not specifying the mixing
probability yet; it will be optimized below. Note also that this
operation is local and could be thought of as having an index
specifying one side of the setup.
\bibitem{NoFundamentalDifficulty} As discussed in
Ref.~\cite{Ashhab}, there is no fundamental difficulty in
implementing this process.
\end{thebibliography}
\end{document}